\documentclass[%
 reprint, 
 superscriptaddress,
 amsmath,amssymb,
 aps,
floatfix,
longbibliography
]{revtex4-2}

\usepackage{orcidlink}

\usepackage{amsmath,amssymb,amsfonts}
\usepackage{algorithmic}
\usepackage{graphicx}
\usepackage{textcomp}
\usepackage{xcolor}
\usepackage{braket}
\usepackage{filecontents,lipsum}

\usepackage{amsthm}
\newtheorem{definition}{Definition}
\newtheorem{protocol}{Protocol}
    
\usepackage{amsmath}
\usepackage{tikz}
\usetikzlibrary{quantikz2,calc,arrows.meta}

\begin{document}

\font\myfont=ptmr at 22pt
\title{Improved Low-Cost Quantum Error Mitigation with \\ Benchmarked-Noise Zero-Noise Extrapolation (bnZNE)}

\author{Joseph Harris \orcidlink{0000-0002-7458-6580}}
\email{joseph.harris@dlr.de}
\author{Kevin Lively \orcidlink{0000-0003-2098-1494}}
\author{Peter K. Schuhmacher \orcidlink{0000-0003-1232-4363}}
\affiliation{Department of High Performance Computing, Institute of Software Technology, German Aerospace Center (DLR),  Rathausallee 12, 53757 Sankt Augustin, Germany}
\date{November 2026}

\begin{abstract}

We theoretically extend and experimentally demonstrate \textit{benchmarked-noise zero-noise extrapolation (bnZNE)}, a simple and low-cost method for quantum error mitigation which aims to outperform the highly popular ZNE method for an equal total shot cost by improving the classical extrapolation step. We recently proposed bnZNE as a new error mitigation method (Harris et al., arXiv:2603.10224 (2026)) and showed that it can mitigate noise to the same order as standard ZNE under gate-wise depolarizing noise. Here, we introduce \textit{randomized bnZNE} which addresses the case of running multiple random circuits per noise level, allowing for the incorporation of noise-tailoring techniques like Pauli twirling and readout error mitigation, and test the method on the IBM and IQM hardware for a plethora of large-scale applications, observing bnZNE to consistently produce more accurate results than standard ZNE. Our approach enhances the extrapolation process to the zero-noise limit by using classically-simulable and structure-mirroring \textit{benchmark circuits}, generated from the standard application circuit, to sensitively measure the effective error rate at each scaled noise level beyond simple integer values. We give a new method for generating tailored benchmark circuits for the IQM superconducting devices and test them experimentally. Our results demonstrate that bnZNE presents a versatile alternative to ZNE which can enable more accurate results at utility scales with an equivalent cost.
\end{abstract}

\maketitle

\section{Introduction}
\label{sec:introduction}

Quantum computing has gained major scientific  and economic interest in recent decades due to the development of algorithms which can, in theory, solve certain problems exponentially faster than classical computers. 
As cloud-based noisy intermediate-scale quantum (NISQ) hardware becomes more readily available, it is inevitable that users must be able to take steps to deal with hardware errors. Quantum error mitigation (QEM) \cite{Strikis2021,Suzuki2022,Takagi2022,Endo2018,Endo2021, filippov2023scalabletensornetworkerrormitigation, RevModPhys.95.045005} is a class of techniques used to mitigate the effects of errors on expectation values calculated with near-term devices by running several modifications of the application circuit and classically post-processing their measurement outcomes. Such methods typically trade accuracy for resource complexity; high-cost methods such as probabilistic error cancellation \cite{vandenBerg2023} or tensor error mitigation \cite{filippov2023scalabletensornetworkerrormitigation,Fischer2026} aim to learn the noise model directly at an  exponentially scaling cost of circuit runs and the requirement for sophisticated post-processing on high-performance computers \cite{Takagi2022, Fischer2026}. Low-cost methods methods can use very few circuits but rely on simplified noise assumptions which break down at larger scales. 

\begin{figure*}[t]
\centering
\resizebox{\textwidth}{!}{%
    \includegraphics[]{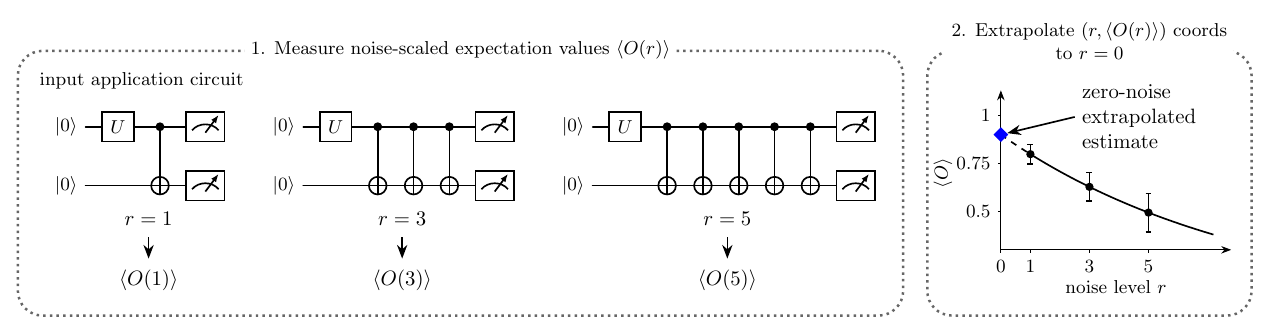}
}
\caption{Overview of standard digital zero-noise extrapolation (ZNE) for quantum error mitigation. A series of noise-scaled circuits are run on the hardware and the expectation value of interest $\braket{O(r)}$ is measured as a function of the noise level $r$, defined as the ratio of two-qubit gates compared to the original circuit. Finally, the datapoints $(r, \braket{O(r)})$ are extrapolated to the zero-noise ($r=0$) limit to obtain an estimate to the true noiseless expectation value.}
\label{fig:zne-overview}
\end{figure*}

In this work, we focus on improving the popular low-cost method of digital zero-noise extrapolation (ZNE) \cite{Giurgica_Tiron_2020, 10313813, PhysRevX.7.021050, PhysRevLett.119.180509, Kandala2019, PhysRevA.102.012426} for expectation value estimation which we illustrate in Figure \ref{fig:zne-overview}. ZNE is conceptually simple and resource efficient but comes at the cost of a bias which increases with circuit depth. In ZNE, the noise of the application circuit is increased via the insertion of identity operations; this is usually with \textit{circuit folding} techniques \cite{Giurgica_Tiron_2020}.
A number of different noise-scaled circuits are then run on hardware and expectation values of interest, $\braket{O}$ are measured as a function $\braket{O(r)}$ of the noise level $r$ for a predetermined set of values $r \in R$. Finally, the datapoints $(r, \braket{O(r)})$ are extrapolated classically to the zero-noise limit $r=0$ to give an estimate $\braket{O(0)}$ to the exact noiseless solution. 

While zero-noise extrapolation remains very popular due to its ease of implementation and low cost, the generally false assumption that $r=0$ corresponds to a noiseless circuit produces a bias which can result in substantial inaccuracies at scale. Factors such as the single-qubit gate noise, temporally changing noise or the plateauing of noise with circuit depth are not fully taken into account \cite{RevModPhys.95.045005}.  

In previous work \cite{harris2026}, we proposed \textit{benchmarked-noise zero-noise extrapolation (bnZNE)} as a heuristically-based alternative to ZNE which uses classically-simulable structure-mirroring \textit{benchmark circuits} to directly measure an effective error rate $\varepsilon(r)$ at each standard ZNE noise level $r$ which are then used instead of $r$ during the extrapolation process. We showed analytically that bnZNE can suppress errors to the same order of magnitude as standard ZNE under gate-wise depolarizing noise and tested it for a simple application on a single hardware platform. In this work, we extend the method with a focus on practical implementations. A new variation to bnZNE, \textit{randomized bnZNE}, allows for integration with other commonly-used error mitigation techniques such as Pauli Twirling \cite{WallmanRandomizedCompiling, Ware_2021, Kim2023}  or readout error mitigation \cite{PhysRevA.105.032620} by addressing the case where multiple randomized circuits are run per ZNE noise level $r$. We give a plethora of new experimental evidence demonstrating that randomized bnZNE consistently outperforms ZNE in practical settings with large-scale circuits, considering two hardware platforms (IBM and IQM) and several quantum applications and subroutines. 

This work is self-contained and is structured as follows. First, we define the benchmark circuit and give several methods for generating benchmark circuits that we use in this work. We then state and discuss our bnZNE and randomized bnZNE methods of error mitigation. We present experimental results from the \texttt{ibm\textunderscore fez} and \texttt{IQM Emerald} devices for a plethora of applications, observing bnZNE to consistently outperform standard ZNE. Finally, we summarize the consequences of our work. Our objective throughout is to provide a self-contained and experimentally-focused resource for those looking to implement bnZNE in their own experiments. 

\section{Defining benchmark circuits}
\label{sec:defining-benchmark-circs}

Here, we define the benchmark circuit which is generated from an input application circuit $\mathcal{A}$. The aim is that these circuits mirror the application circuit's gate structure and thus are subjected to comparable overall noise. This definition is simplified compared to the original definition in \cite{harris2026}, although the two are logically equivalent. 

\begin{definition}[Benchmark circuit]
    Given as input an application circuit $\mathcal{A}$ and Pauli string observable $O$, we define a \textit{benchmark circuit} $\mathcal{B}(\mathcal{A})$ to be any circuit satisfying the following three constraints:
    \begin{itemize}
    \item[C1:] Circuits $\mathcal{A}$ and $\mathcal{B}$ may be transpiled to the target hardware such that they contain the same sequence of native gates and measurements, up to replacing existing gates with new equivalent-error gates.
    \item[C2:] The measurement outcome of $\mathcal{B}$ is a bitstring $b = (b_q)_{q \in Q(O)}$ which can be efficiently classically determined, where $Q(O) = \{q: O_q \neq I\}$ is the set of qubits on which the input observable acts non-trivially. 
    \end{itemize}
    \label{def:bm}
\end{definition}

Without loss of generality, we assume that all circuit measurements are performed with respect to the standard $Z$ basis. We hence assume that $O \in \{I,Z\}^{\otimes n}$, i.e.\ any basis changes are included within the circuit. Note that this also implies $\braket{O}_\mathcal{B} = \pm 1$ with the sign depending on the bitstring $b$.

Definition \ref{def:bm} is broad enough that it is possible to heavily tailor the method for generating the benchmark circuit depending on the scenario. We now present several methods for generating benchmark circuits; the first are tailored to particular hardware such as the IBM and IQM devices whilst the latter is tailored to repeated-layer circuits.

\subsection{Hardware-tailored benchmark circuits}
\label{sec:ibm-benchmark-circs}

Ideally, benchmark circuits should be generated directly from the hardware-compiled application circuit in order to work with the least noisy (i.e.\ most efficiently compiled) representation of the application circuit as possible. Here, we give a general method of generating benchmark circuits which satisfy this condition and give two specific examples for the IBM and IQM hardware. 

We firstly assume that the hardware's native gate set contains one two-qubit entangling gate; the generalisation to many such gates is straightforward. A set of general, sufficient conditions on the hardware for the ability to generate hardware-tailored benchmark circuits is as follows. 
\begin{enumerate}
    \item There exists an orthonormal single-qubit basis set $\{\ket{\psi_1}, \ket{\psi_2}\}$ such that the native two-qubit gate maps tensor products of these states to tensor products, i.e. for all $i,j \in \{1,2\}$ there exist $k, l \in \{1,2\}$ such that the native two-qubit gate maps $\ket{\psi_i}\otimes\ket{\psi_j} \mapsto \ket{\psi_k}\otimes\ket{\psi_l}$ up to an irrelevant global phase. 
    \item Each single-qubit gate can be replaced with another single-qubit gate of equivalent error (e.g.\ by changing an angle parameter or replacing the gate entirely) such that the new gate acts on the states $\{\ket{\psi_1}, \ket{\psi_2}\}$ either as identity or exchanging $\ket{\psi_1} \leftrightarrow \ket{\psi_2}$, again up to a global phase.
\end{enumerate}
On the surface, these conditions seem restrictive on the kinds of hardware that we may apply this method to. In practice, however, they cover many platforms including those we consider in this work. To generate a benchmark circuit, we may then apply the following procedure:

\begin{protocol}[Hardware-agnostic benchmark circuits]
Given a circuit to be run on hardware satisfying the conditions above, we apply the following steps:
\begin{enumerate}
    \item Map the input $\ket{00\cdots0}$ into the $\{\ket{\psi_1}, \ket{\psi_2}\}$ basis; this can in general be done using a small number of single-qubit gates per qubit \cite{PhysRevA.52.3457}.
    \item Apply the application circuit with gate replacements according to point 2 above.
    \item For each qubit $q$ on which $O$ acts non-trivially, map the state $\ket{\psi_i}$ into a computational basis state $\ket{i}$, $i=0$ or $1$, using a small number of single-qubit gates. 
\end{enumerate}
\end{protocol}

Note that the state of such a circuit remains a product state throughout. Thus, we can efficiently keep track of the full circuit state, which is either a bitstring (at the beginning and end) or takes the form $\bigotimes_{q} \ket{\psi_{i_q}}$. We now give two examples of benchmark circuits tailored for the IBM and IQM devices respectively. Conveniently, in both cases we have $\{\ket{\psi_1}, \ket{\psi_2}\} = \{\ket{0},\ket{1}\}$. 

\textit{Benchmark circuits for IBM hardware.} In this work, we will use the \textit{Heron r2} IBM superconducting quantum computers and thus we give a particular version of the above method tailored to this platform.
These devices use $CZ$ as the native two-qubit gate, $R_z(\theta), \sqrt{X}, X$ as the native single-qubit gates and measure in the standard $Z$-basis. The $\sqrt{X}$ and $X$ gates are implemented via equal-time pulses, such that they are reported to have identical error rates \cite{ibm_quantum}, whilst the $R_z$ gate is simulated virtually for arbitrary angle by adjusting the relative phase between the projection of the qubit vector on the Bloch sphere and the microwave control tone and can be viewed as error-free \cite{PhysRevA.96.022330}. Note that the two-qubit $CZ$ gate and single qubit $R_Z(\theta)$ and $X$ gates act trivially on tensor products of the computational basis states $\{\ket{0}, \ket{1}\}$, applying at most single bit-flips or an ignorable global phase. Thus, we generate benchmark circuits from the compiled application circuits as follows: 
\begin{protocol}[Benchmark circuits for IBM Heron r2 devices]
Given as input an application compiled in terms of the native gate set $\{CZ, R_Z(\theta), \sqrt{X}, X\}$, the corresponding benchmark circuit may be generated by replacing each $\sqrt{X}$ gate with an $X$ gate.
\label{protocol:ibm}
\end{protocol}
The resulting circuit acts trivially on the input all-zero state, with the $X$ gates applying single bit-flips and the $CZ$ and $R_Z(\theta)$ gates acting as identity. Thus, the output state is a classical bitstring that can be efficiently computed (condition C2). 
Finally, the compiled application and benchmark circuits have precisely the same gate layout, with the only difference being $\sqrt{X}$ gates replaced by equivalent-error $X$ gates (condition C1).

\textit{Benchmark circuits for IQM hardware.} We also use the 54-qubit IQM Emerald superconducting device with native gate set $CZ, R(\theta, \phi) = e^{-i\frac{\theta}{2}\left(\cos\phi X + \sin\phi Y\right)}$. The latter gate corresponds to a rotation $\theta$ around the $\cos\phi X + \sin\phi Y$ axis of the Bloch sphere. We note in particular that the gates $X = R(\pi,0)$ and $Y = R(\pi, \pi/2)$ both act  on a computational basis state by introducing at most bit flips and global phases. We hence propose the following method of generating a (random) IQM-tailored benchmark circuit from an application circuit expressed in terms of $CZ$ and $R$ gates: 
\begin{protocol}[Benchmark circuits for IQM hardware]
    Given as input an application circuit compiled in terms of the native gate set $\{CZ, R(\theta,\phi)\}$, a random corresponding benchmark circuit may be generated as follows: for each $R(\theta, \phi)$ gate in the application circuit, replace it with either an $X = R(\pi,0)$ or $Y = R(\pi, \pi/2)$ gate, chosen with probabilities $\cos^2\phi, \ \sin^2\phi$ respectively.
    \label{protocol:iqm}
\end{protocol}
At the time of writing, each available IQM device has this native gate set, meaning our method can be applied to any IQM device. 

Like before, the resulting benchmark circuit has a single bitstring output which can be efficiently calculated. This method takes into account that the noise of each $R(\theta, \phi)$ gate may depend on the axis of rotation by sampling from both the $X$ and $Y$ axes independently for each gate using a probability distribution based on the original gate angle. Unlike in the IBM case, this means that a family of random benchmark circuits can be generated from a single application circuit. 

We note here that the benchmark circuits introduced in this section are non-entangling; the circuit state remains a product state throughout. Thus, the circuits could be less prone to the long-range crosstalk errors that are typical of many real devices and may affect the application circuit. To this end, we also give a second method which generates entangling benchmark circuits and is tailored to circuits with repeating layers.

\subsection{Entangling benchmark circuits for layered applications}
\label{sec:entangling-benchmark-circs}

We give an additional method for generating benchmark circuits which is tailored to Floquet circuits, i.e.\ circuits consisting of multiple layers consisting of the same native gate structure. The main benefit to this method is that the corresponding circuits are entangling, and thus we expect them to better capture long-range noise behaviour in the hardware compared to the previous method.

We assume as input an application circuit such that:
\begin{itemize}
    \item The application consists of $2L$ layers where each layer contains an identical layout of single and two-qubit gates.
    \item Each layer can be inverted to produce a new layer containing the same consistent layout of single and two-qubit gates, up to phase and angle changes.
\end{itemize}
This method is specifically targeted to Trotterized circuits -- commonly used for quantum simulation applications -- where each layer may correspond to a series of Pauli rotations which can be straightforwardly inverted. We also note that one can assume $L=1$ without loss of generality. 
\begin{protocol}[Benchmark circuits for layered applications]
    Given as input a layered application circuit satisfying the assumptions above, the corresponding benchmark circuit may be generated as follows:
    \begin{enumerate}
    \item Let the input application circuit be 
        \begin{equation}
            \mathcal{A} = U_{2L} U_{2L-1} \cdots U_1
        \end{equation}
        where each $U_i$ is a single layer obeying the properties above.
    \item Define the corresponding benchmark circuit to be
    \begin{equation}
        \mathcal{B} = \left( U_1^{-1} U_2^{-1} \cdots U_L^{-1} \right) \left( U_L U_{L-1} \cdots U_1 \right).
    \end{equation}
\end{enumerate}
\label{protocol:entangling}
\end{protocol}
In other words, we apply the first $L$ layers of the application circuit followed by their inverses to produce a circuit which is logically equivalent to identity, i.e.\ it has classical output string $0^n$ (condition C3) and, since $O \in \{I,Z\}^{\otimes n}$, we have $\braket{O}_\mathcal{B}^\text{exact} = 1$ (condition C2). By the second assumption above, the application and benchmark circuit have identical structure (condition C1). 

\section{Benchmarked-noise zero-noise extrapolation (bnZNE)}

\begin{figure*}[t]
\centering
\resizebox{\textwidth}{!}{%
    \includegraphics[]{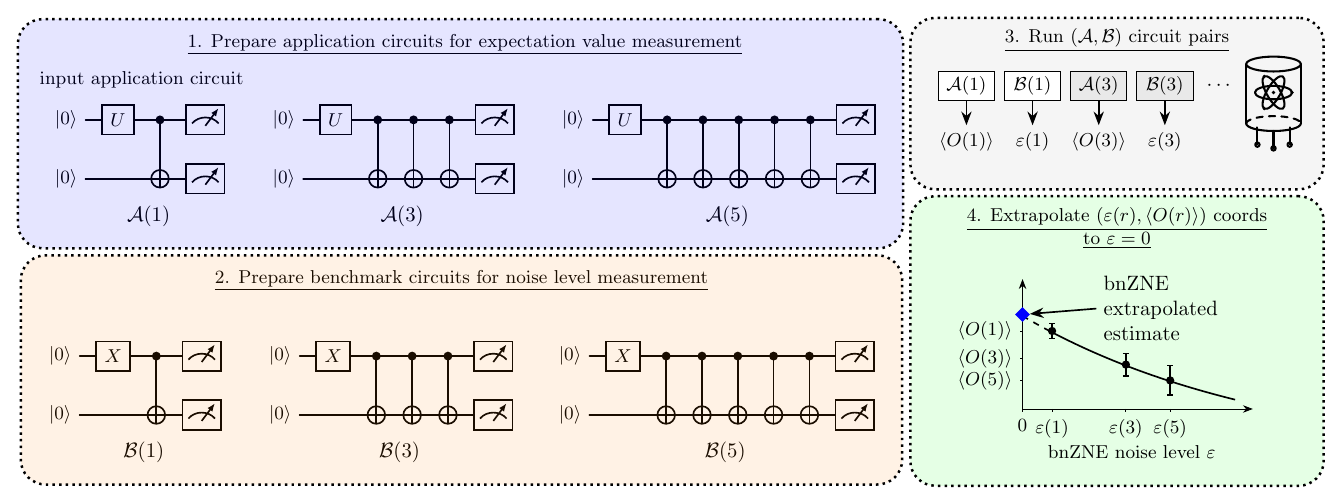}
}
\caption{Overview of our benchmarked-noise zero-noise extrapolation (bnZNE) method: 1. Noise-scaled application circuits are produced using standard gate folding techniques. 2. A benchmark circuit is generated for each noise-scaled application circuit using a chosen generation method (Section \ref{sec:defining-benchmark-circs}). 3. The application, benchmark circuit pairs are run consecutively on the hardware: each application circuit measures the expectation value of interest $\braket{O(r)}$ and each benchmark circuit measures the corresponding bnZNE noise level $\varepsilon(r)$. 4. The datapoints  $(\varepsilon(r), \braket{O(r)})$ are extrapolated to the zero-noise ($\varepsilon=0$) limit to obtain an estimate to the true noiseless expectation value. We choose the folding scale factors $r=1,3,5$ but of course other values can be chosen so as to appropriately sample the bnZNE noise level space $\varepsilon \in [0,1]$.}
\label{fig:bnzne-overview}
\end{figure*}

Standard zero-noise extrapolation (ZNE) (Figure \ref{fig:zne-overview}) aims to construct an error-mitigated expectation value $\braket{O}_\mathcal{A}^\text{QEM}$ of some observable $O$ with respect to an application circuit $\mathcal{A}$ via the simple protocol given below   \cite{Giurgica_Tiron_2020, 10313813, PhysRevX.7.021050, PhysRevLett.119.180509, Kandala2019, PhysRevA.102.012426}.
\begin{protocol}[Standard zero-noise extrapolation (ZNE)]
\label{protocol:zne}
Given as input an application circuit $\mathcal{A}$ and Pauli string observable $O$ acting non-trivially on some subset of qubits $Q \equiv \{i : O_i \neq I\}$ whose expectation value is to be measured:
\begin{enumerate}
    \item Generate a series of noise-inflated circuits $\mathcal{A}(r)$ for a series of pre-determined noise levels $r \in R$ (often $R = \{1,3,5\}$).
    \item Run the noise-scaled circuits on the hardware and measure the corresponding noise-inflated expectation values $\braket{O(r)}_\mathcal{A}$.
    \item Extrapolate the datapoints $(r, \braket{O(r)}_\mathcal{A})$ to the zero-noise limit $r=0$, producing an estimate $\braket{O}_\mathcal{A}^\text{QEM} \equiv \braket{O(0)}_\mathcal{A} \approx \braket{O}_\mathcal{A}^\text{exact}$.
\end{enumerate}
\end{protocol}

The noise-inflated circuits are produced using a fixed method of \textit{folding}; \textit{local folding} interleaves individual gates and their inverses while \textit{global folding} applies the whole circuit followed by its inverse \cite{Giurgica_Tiron_2020}. As discussed in the introduction, the accuracy of ZNE performed with simple $r \rightarrow 0$ extrapolation relies on an assumed scaling relationship between the inferred ZNE noise level $r$ and the true noise of the implemented circuit; in particular that the extrapolated $r=0$ expectation value corresponds to one obtained from a circuit which is noise-free. Factors such as the single-qubit gate noise, temporally changing noise or the plateauing of noise with circuit depth are not fully taken into account, leading to inaccuracies at larger scales which we also observe experimentally in Section \ref{sec:results}.

Our modification to standard ZNE aims to improve the classical extrapolation process by better contextualising the relative `noise levels' between the noise-inflated circuits beyond the simple odd-integer description. We do this by running benchmark circuits at each ZNE noise level and using them to calculate an inferred noise level which may be any real number between 0 and 1; hence we call our method \textit{benchmarked-noise zero-noise extrapolation (bnZNE)}. A pictorial overview of the method is given in Figure \ref{fig:bnzne-overview}. 

The method is then as follows. 
\begin{protocol}[Benchmarked-noise zero-noise extrapolation (bnZNE)]
\label{protocol:bnzne}
Given as input an application circuit $\mathcal{A}$ and Pauli string observable $O$ acting non-trivially on some subset of qubits $Q \equiv \{i : O_i \neq I\}$ whose expectation value is to be measured:
\begin{enumerate}
    \item Generate the benchmark circuit $\mathcal{B}(\mathcal{A})$ using a chosen method. By condition C2 (Section \ref{sec:defining-benchmark-circs}), the measurement outcome of $\mathcal{B}$ is a single efficiently-determinable bitstring $b = (b_q)_{q \in Q}$.
    \item Run both $\mathcal{A}$ and $\mathcal{B}$ on the device at each of the chosen noise levels $r \in R$, ensuring to run each benchmark circuit immediately after its corresponding application circuit. For $\mathcal{A}$, calculate and store the noise-scaled expectation values $\braket{O(r)}_\mathcal{A}$. For $\mathcal{B}$, use the measurement outcomes to calculate the ensemble of probabilities $p(i;q,r)$ of obtaining measurement outcome $i \in \{0,1\}$ on qubit $q \in Q$ when running the noisy benchmark circuit $\mathcal{B}(r)$ at ZNE noise level $r \in R$,
    \begin{equation}
    \begin{split}
        p(i;q,r) &=  \text{Tr}_{\bar{q}}\left[\ket{i}\bra{i}_q U_{\mathcal{B}(r)}\rho_0U_{\mathcal{B}(r)}^{\dagger}\right] \\
        &= \sum_j \bra{j}_{\bar{q}} \bra{i}_q \left( U_{\mathcal{B}(r)}\rho_0U_{\mathcal{B}(r)}^{\dagger} \right) \ket{i}_q \ket{j}_{\bar{q}},
    \end{split}
    \end{equation}
    where $\bar{q}$ denotes all qubits except for $q$ and $\rho_0$ is the initial all-zero state.
    \item Calculate the associated bnZNE noise level at ZNE noise level $r$, which we define to be
    \begin{equation}
        \varepsilon(r) \equiv \prod_{q \in \mathcal{Q}} (1-p(b_q; q,r)),
        \label{eq:bnzne-error-rate}
    \end{equation}
    i.e.\ the product of the probabilities of obtaining the wrong measurement outcome across each of the measured qubits. 
    \item Extrapolate the datapoints $\left(\varepsilon(r), \braket{O(r)}_\mathcal{A}\right)$ to noise level $\varepsilon = 0$ to reconstruct the corresponding noiseless value.
\end{enumerate}
\label{protocol:bnzne}
\end{protocol}

In \cite{harris2026}, we prove that bnZNE can mitigate errors to at least the same order of magnitude as standard ZNE under gate-wise depolarizing noise. As we will show later, in experimental settings we consistently found our method to outperform ZNE. Although bnZNE requires running twice as many circuits as ZNE, we tested the method using an equal total number of shots by using half as many shots per bnZNE circuit. Hence, our method can provide a practical accuracy advantage with no additional overhead in quantum cost.

At this point, we draw attention to three limitations to the above protocol. Firstly, Protocol \ref{protocol:bnzne} cannot be combined with other methods for error mitigation and noise-tailoring such as Pauli twirling (PT) or twirled readout error extinction (TREX), each of which relies on running several random circuits per ZNE noise level and averaging over their results rather than running just one. PT tailors coherent errors into stochastic Pauli channels, which generally exhibit well-behaved scaling with standard ZNE \cite{Cai2019} while TREX mitigates readout errors by diagonalizing each qubit's readout error channel. Each of these methods applies some random process $P$ to a given circuit $\mathcal{C}$, replacing it by a set of $N$ random circuits $\mathcal{C}_1, \mathcal{C}_2, \dots, \mathcal{C}_N$ which are logically equivalent to $\mathcal{C}$ but may contain additional gates. Measured expectation values are then averaged over each of the $N$ random circuits; for PT, this averaging results in effective Pauli noise while for TREX this allows for measurement error to be later cancelled out with classical post-processing. 

Secondly, Protocol \ref{protocol:bnzne} only uses a single benchmark circuit which is fixed throughout the protocol and to which noise scaling is applied. However, in cases where the benchmark circuits are generated randomly (e.g.\ in the IQM case, Protocol \ref{protocol:iqm}), it is more instructive to generate several random benchmark circuits and average over their results.

Finally, the original bnZNE protocol only samples $\varepsilon$ once for each value of $r$. In the realistic case of temporally varying noise, random variations in measurements could penalize the quality of the extrapolation process if only a small number of measurements are taken. Instead, it would be better to measure $\varepsilon$ multiple times per fixed $r$ such that we could sample multiple values across the $\varepsilon \in [0,1]$ range. 

To this end, we introduce  \textit{randomized bnZNE} which addresses each of the limitations above by allowing for independent random processes to be applied to the application and benchmark circuits and for a greater overhead of circuit runs. We state the protocol below and interpret it after. 

\begin{protocol}[Randomized benchmarked-noise zero-noise extrapolation (randomized bnZNE)]
\label{protocol:randomized-bnzne}
Given as input an application circuit $\mathcal{A}$ and Pauli string observable
$O$ acting non-trivially on some subset of qubits
$Q \equiv \{i : O_i \neq I\}$:
\begin{enumerate}
    \item Choose a randomisation process $P$ (e.g.\ Pauli twirling or TREX)
    and a number $N$ of randomisations. Apply $P$ to the application circuit
    $\mathcal{A}$ to generate $N$ randomized application circuits
    \[
        \mathcal{A}_n = P_n(\mathcal{A}),
        \qquad n=1,\ldots,N,
    \]
    where $P_n$ denotes an independently sampled instance of the
    randomisation process $P$.

    \item Choose a benchmark generation process $Q$ and use it to generate
    a corresponding benchmark circuit $\mathcal{B}_n$ from each randomized
    application circuit $\mathcal{A}_n$:
    \[
        \mathcal{B}_n = Q_n(\mathcal{A}_n),
        \qquad n=1,\ldots,N.
    \]
    By condition C2 (Section \ref{sec:defining-benchmark-circs}), the
    measurement outcome of each benchmark circuit is a single efficiently-
    determinable bitstring $b^{(n)}=(b_q^{(n)})_{q\in Q}$.

    \item For each chosen ZNE noise level $r\in R$, generate the corresponding
    noise-scaled application and benchmark circuits
    \[
        \mathcal{A}_n(r),\qquad \mathcal{B}_n(r),
        \qquad n=1,\ldots,N.
    \]

    \item For each $r\in R$ and each $n\in\{1,\ldots,N\}$, run the corresponding
    application and benchmark circuits $\mathcal{A}_n(r)$ and $\mathcal{B}_n(r)$,
    ensuring to run each benchmark circuit immediately after its corresponding
    application circuit. For the application circuit, calculate and store the
    noisy expectation value
    \[
        \braket{O(r)}_{\mathcal{A}_n}.
    \]
    For the corresponding benchmark circuit, use the measurement outcomes to
    calculate the probabilities
    \[
        p_n(i;q,r)
        =
        \mathrm{Tr}_{\bar q}
        \left[
        \ket{i}\bra{i}_q
        U_{\mathcal{B}_n(r)}
        \rho_0
        U_{\mathcal{B}_n(r)}^\dagger
        \right],
    \]
    where $\bar q$ denotes all qubits except for $q$ and $\rho_0$ is the
    initial all-zero state. From these probabilities, calculate each
    corresponding bnZNE noise level
    \[
        \varepsilon_n(r)
        \equiv
        \prod_{q\in Q}
        \left(1-p_n(b_q^{(n)};q,r)\right).
    \]

    \item For each ZNE noise level $r$, average the application expectation
    values and associated bnZNE noise levels over the $N$ randomisations:
    \[
        \overline{\braket{O(r)}_{\mathcal{A}}}
        =
        \frac{1}{N}
        \sum_{n=1}^{N}
        \braket{O(r)}_{\mathcal{A}_n},
    \]
    \[
        \overline{\varepsilon(r)}
        =
        \frac{1}{N}
        \sum_{n=1}^{N}
        \varepsilon_n(r).
    \]

    \item Extrapolate the datapoints
    \[
        \left(
        \overline{\varepsilon(r)},
        \overline{\braket{O(r)}_{\mathcal{A}}}
        \right),
        \qquad r\in R,
    \]
    to the zero-noise limit $\varepsilon=0$ to reconstruct the corresponding noiseless value.
\end{enumerate}
\end{protocol}

Protocol \ref{protocol:randomized-bnzne} amounts to a generalization of the original bnZNE protocol: if $P = Q = I$ and $N=1$ then we recover Protocol \ref{protocol:bnzne}. However, this also allows for additional cases: $P \neq I$ allows for the incorporation of a noise-tailoring technique like PT or TREX and $Q \neq I$ allows for a benchmark generation method which produces random circuits and our protocol allows for the two to be combined. $N>1$ mitigates the effect of temporal flucations in the measured values by averaging over multiple runs at the same noise level.

We employ randomized ZNE for our experimental results in Section \ref{sec:results}, where our experiments on IBM combine bnZNE with Pauli twirling and twirled readout error extinction for readout error mitigation and our experiments on IQM incorporate the random tailored benchmark circuits produced using Protocol \ref{protocol:iqm}.   

There exist other methods which modify standard ZNE in order to improve the extrapolation procedure. One, \textit{inverted-circuit ZNE (IC-ZNE)} \cite{PhysRevA.110.042625}, gauges the noise-level by applying the noise-scaled circuit followed by its inverse, and measuring the probability of the all-zero state. Another considers circuits with repeating layers and benchmarks the fidelity of a single-layer \cite{kim2025enhancedextrapolationbasedquantumerror}. Our method is more generalized, makes no assumptions on the circuit structure when using the hardware-tailored benchmark circuits and requiring no overhead of circuit depth compared to the original circuit beyond standard ZNE. Moreover, our method has the potential to be tailored to particular algorithms or hardware platforms by varying the benchmark circuit generation method beyond the two examples we give in this work.

\section{Results}
\label{sec:results}

\begin{figure*}
    \centering
    \resizebox{\textwidth}{!}{%
    \includegraphics[]{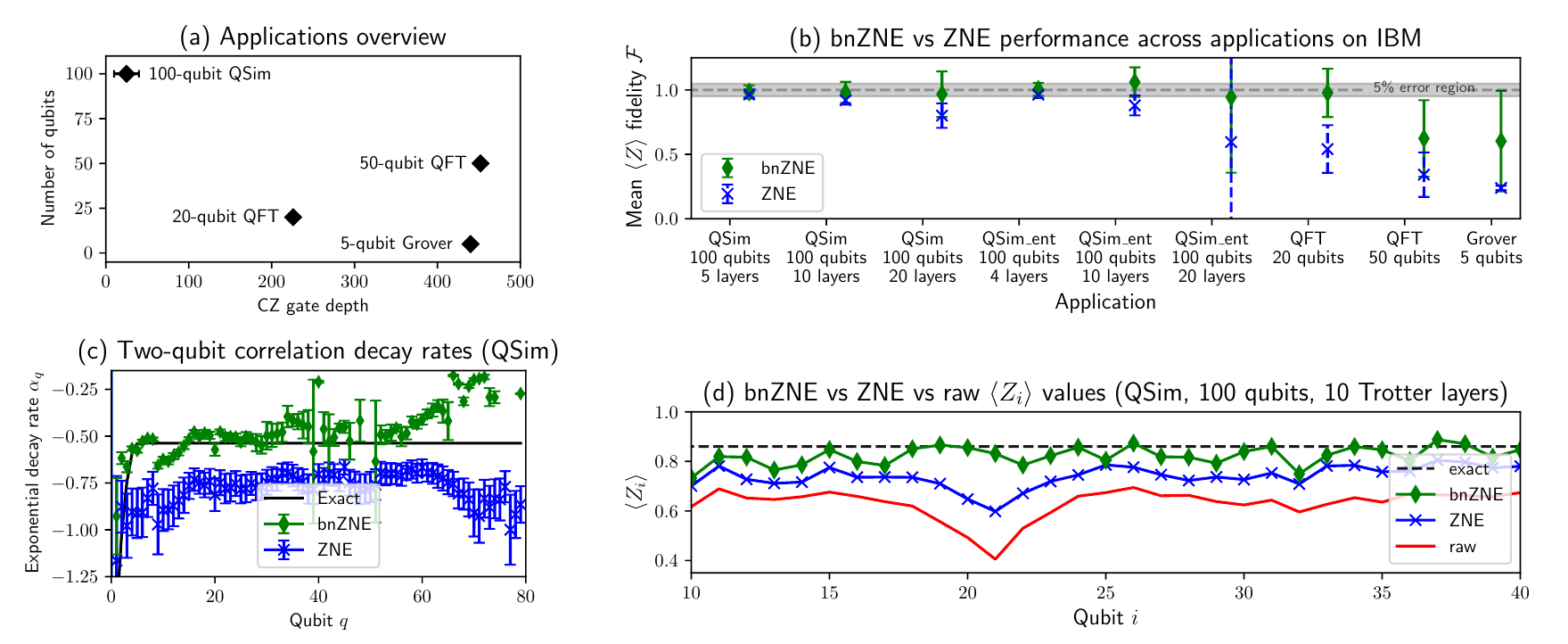}
    }
    \caption{Overview of experimental results obtained from the 156-qubit \texttt{ibm\textunderscore fez} superconducting device. (a) gives an overview of the applications implemented on the quantum hardware in terms of their two-qubit gate depth and the total number of qubits. (b) compares the mean single-qubit $\braket{Z}$ fidelities obtained for bnZNE and ZNE across each application. (c) compares bnZNE and ZNE for the calculation of two-qubit correlation decay rates for the case of 100 qubit, 20 layer Trotterized circuit (equation (\ref{eq:ki-circ})) simulating the time evolution of a kicked Ising Hamiltonian. (d) compares the bnZNE, ZNE and raw single-qubit $\braket{Z_i}$ expectation values as a function of the qubit index $i$ for 100 qubit, 10 Trotter layer quantum simulation. Across all experimental results, we observe bnZNE to consistently produce more accurate results than standard ZNE.}
    \label{fig:ibm-combined}
\end{figure*}

\begin{figure*}
    \resizebox{0.9\textwidth}{!}{%
    \includegraphics[]{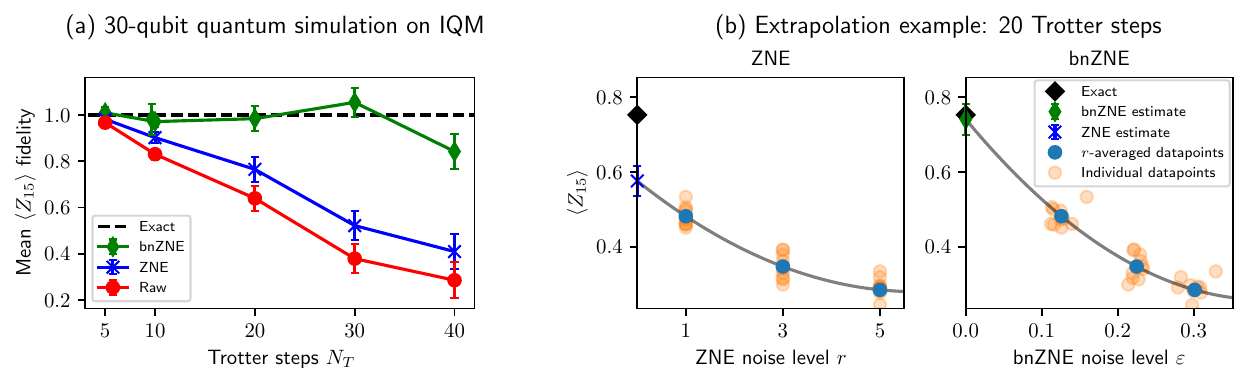}
    }
    \caption{Overview of experimental results obtained from the 54-qubit \texttt{IQM Emerald} superconducting device. In each case we run a quantum circuit (equation (\ref{eq:ki-circ})) simulating the Trotterized time evolution of a 30-qubit one-dimensional kicked Ising Hamiltonian. (a) compares the mean fidelity of a single-qubit expectation value measurement using bnZNE, ZNE and no error mitigation as a function of the number of Trotter steps $N_T$ in the circuit. We observe a clear advantage using bnZNE compared to standard ZNE. (b) gives an example of the extrapolation for ZNE and bnZNE, where in each case the three datapoints are fitted to a quadratic function. The measured bnZNE noise levels accurately contextualize the relative noise between different $r$ values, resulting in an greatly improved extrapolation step.}
    \label{fig:iqm-combined}
\end{figure*}

\subsection{Methods and algorithms used}
\label{sec:methods-and-algorithms}

We explore the efficacy of randomized bnZNE (Protocol \ref{protocol:randomized-bnzne})  against standard ZNE (Protocol \ref{protocol:zne} for four different application circuits with varying numbers of qubits and gate depths. In this section, for simplicity we refer to randomized bnZNE simply as bnZNE. Each application was and run on the 156-qubit \texttt{ibm\textunderscore fez} superconducting device which uses a heavy-hexagon qubit layout and has the basis gate set $\{CZ, R_Z(\theta), \sqrt{X}, X\}$. \cite{ibm_quantum}. We also performed additional experiments for one application using the 54-qubit \texttt{IQM Emerald} superconducting device, which has a standard grid layout and the basis gate set $\{CZ, R(\theta, \phi)\}$  \cite{iqmemerald2025}.

To perform ZNE, for each application we generate noise-scaled application circuits for the three standard ZNE noise levels $\{1,3,5\}$ using local circuit folding on the two-qubit $CZ$ gates; for each circuit we then perform 32 instances of Pauli twirling (PT) \cite{WallmanRandomizedCompiling, Ware_2021, Kim2023}  as well as dynamical decoupling (DD) \cite{Ezzell2023} and twirled readout error extinction (TREX) \cite{PhysRevA.105.032620} to improve noise resilience. For the extrapolation process, we use exponential fits of the 96 total datapoints (32 per standard ZNE level). Each circuit is run with 2048 shots for a total of 196,608 shots per application. 

For bnZNE, we use Protocol \ref{protocol:randomized-bnzne}.  with $P$ the combined random process applying PT, DD and TREX to generate $N=32$ random circuits and $Q$ the chosen benchmark generation process which varies depending on the application and hardware platform. For the IBM device, we use Protocol \ref{protocol:ibm} for each application as well as  \ref{protocol:entangling} for one application (see below). For the IQM device we use Protocol \ref{protocol:iqm}. Each of the $2N$ (application and benchmark) circuits are run with 1024 shots for a total of 196,608 shots per application, the same as for ZNE.

In each case, we compare measured expectation values against the exact values calculated either directly in Qiskit using the EstimatorV2 primitive \cite{qiskit2024} or via tensor network simulations using ITensors.jl \cite{itensor}. We now describe the individual applications used in our experiments.

\textbf{QSim:}
We consider the quantum simulation of a one-dimensional kicked Ising chain, using 100 qubits on IBM and 30 qubits on IQM. The application circuit is a Trotterized layered circuit,
\begin{equation}
    \mathcal{A} = [L_2L_1]^{N_T},
    \label{eq:ki-circ}
\end{equation}
where 
\begin{equation}
    L_1 = \prod_{i} R_{X_i} \left(\theta_1 \right),  \quad  L_2 = \prod_{\braket{i,j}} R_{Z_i Z_j}\left( \theta_2 \right)
\end{equation}
with the first sum over all qubits and the second over each nearest-neighbour pair. We set $\theta_1 = -\pi/8, \  \theta_2 = -\pi/2$ and choose values of $N_T$ between 4 and 40 Trotter layers. We also considered this application in our previous work \cite{harris2026} but present new experimental results here. On the IBM hardware, in addition to our IBM-tailored benchmark circuits (Protocol \ref{protocol:ibm}) we also run separate experiments using our benchmark circuits for layered applications (Protocol \ref{protocol:entangling}) and refer to this as \textbf{QSim\textunderscore ent}. In the IBM case the deepest base application circuit for 20 Trotter layers contains 1980 CZ gates with a CZ-depth of 40 (2 per Trotter layer) when compiled to the \texttt{ibm\textunderscore fez} device and 1160 CZ gates with a CZ-depth of 80 for the \texttt{IQM Emerald} device.

The remainder of the applications were only run on the IBM hardware due to access limitations. 

\textbf{QFT:} We consider two circuits which prepare the state $\ket{+}^{\otimes N}$ and implement a constrained quantum Fourier transform \cite{Barros2026} to produce the all-zero state. 
The standard quantum Fourier transform implements the mapping
\begin{equation}
    \ket{x} \mapsto \sum_{k=0}^{N-1} \omega_N^{xk} \ket{k}
\end{equation}
on basis states $\ket{x} \in \{0,1,\dots,N-1\}$, where $\omega_N = e^{2\pi i/N}$ is an $N$-th root of unity. 
The first application circuit uses 20 qubits and, to reduce the number of SWAPS, restricts the controlled operations to qubits within a distance 6 of each other, resulting in 374 CZ gates and a CZ-depth of 226. The latter uses 50 qubits and restricts operations to a distance of 4, resulting in 795 CZ gates and a CZ depth of 452. These represent relatively large and deep algorithms, with the high depth caused by the ladder structure of the algorithm. In this case, we used the IBM-tailored benchmark circuits (Protocol \ref{protocol:ibm}). 

\textbf{Grover:} Finally, we consider a 5-qubit Grover search circuit \cite{grover}, implementing an unstructured search for a randomly chosen classical bitstring of length 5. The application circuit contains 795 CZ gates with a CZ-depth of 425, thus representing a `short but deep' circuit. In this case, we used the IBM-tailored benchmark circuit design from Section \ref{sec:ibm-benchmark-circs}, Again we use the IBM-tailored benchmark circuits (Protocol \ref{protocol:ibm}), although it may be possible to generate entangling benchmark circuits when the number of Grover steps is even.

An overview of these algorithms -- in terms of the total number of qubits and CZ-depth on the IBM hardware -- is shown in Figure \ref{fig:ibm-combined}(a). Our intention with these choices is to cover a wide-ranging variety of circuit sizes and use-cases, and in particular to explore large, \textit{utility-scale} instances. Despite their relative shallow depth, quantum simulation algorithms (in particular kicked Ising simulations) have been recently well-studied in the context of demonstrating quantum advantage \cite{Kim2023, PRXQuantum.5.010308, doi:10.1126/sciadv.adk4321}, whilst the Grover search and quantum Fourier transform are both deep-circuit quantum subroutines with links to quantum advantage claims \cite{Lewis2025}. Hence we seek not only to demonstrate the usefulness of our method but also the method's current limitations for our particular choices of hardware.

\subsection{Experimental results}

We start by discussing our experimental results obtained from the IBM device which are given in Figure \ref{fig:ibm-combined}. In Figure \ref{fig:ibm-combined}(b), we consider the measurement of single-qubit $\braket{Z}$ expectation values across each application. Specifically, we calculate the average fidelity,
\begin{equation}
    \text{mean fidelity } \mathcal{F} = \frac{1}{N} \sum_{q=1}^{N} \frac{\braket{Z_q}_\mathcal{A}^\text{QEM}}{\braket{Z_q}_\mathcal{A}^\text{exact}}
\end{equation}
for each QEM method (ZNE and bnZNE). Each application uses the IBM-tailored benchmark circuits (Protocol \ref{protocol:ibm}), except \textit{QSim\textunderscore ent} which uses benchmark circuits for repeated-layer applications (Protocol \ref{protocol:entangling}). For both ZNE and bnZNE we use exponential fits for the extrapolation. Our results show improvements to the mean fidelity over standard ZNE for every application we consider, often with a mean error under 5\%, as indicated by the shaded region. We also observe a variance that increases for bnZNE with the complexity of the circuit, an example of the general bias-variance trade-off observed amongst error mitigation protocols \cite{RevModPhys.95.045005}. Consistently we find that deeper circuits present a greater challenge for bnZNE and ZNE to recover a signal from the underlying noise. Nonetheless bnZNE's mean expectation value fidelities consistently out-perform ZNE, albeit with at the cost of a large variance. This is particularly notable in the case of the Grover application whose depth, despite only being on 5 qubits, results in normal ZNE performing very poorly. 

Notably this trend contradicts the normal order-of-magnitude scaling arguments for error mitigation, which asserts that the principle quantity indicating the strength of noise in the system, and thus the amount of overhead required for mitigation, is the `circuit area' $NL$ for $N$ qubit circuits at two-gate depth $L$  \cite{filippov2024scalabilityquantumerrormitigation,zimborás2025mythsquantumcomputationfault}. For a given noise level $\epsilon$, the overhead of QEM is then asserted to scale as $\sim\text{exp}(\epsilon NL)$. Each circuit application here is run with the same amount of overhead, i.e. TREX, Twirl and $r$-scaled two-qubit gates -- with bnZNE adding an additional factor of two for the benchmark circuits. Despite this, we note that the bnZNE fidelities obtained for the 5-qubit Grover and 50-qubit QFT are very similar, despite having very different circuit areas of 2,125 and 22,600 respectively. This seems to indicate that in fact that the depth of the circuit plays a stronger role than this simple analysis would suggest, at least for these QEM methods. 

In Figure \ref{fig:ibm-combined}(d) we compare the first fifty $\braket{Z}$ expectation values obtained for the QSim\textunderscore ent application using ZNE and bnZNE to the exact and raw (unmitigated $r=1$) values. We see that, at the cost of increased variance, bnZNE clearly improves on the residual mitigation bias in standard ZNE, even in cases where the signal-to-noise ratio -- indicated via the raw results -- is quite low.

Finally, in Figure \ref{fig:ibm-combined}(c) we consider the calculation of multi-qubit observables. As a physically-motivated example, we consider the QSim application which simulates the time evolution of an Ising chain. We define the correlation between two qubits in the chain to be
\begin{equation}
    \braket{Z_iZ_j}^\text{corr} \equiv \braket{Z_iZ_j}-\braket{Z_i}\braket{Z_j}.
\end{equation}
In general, we expect correlations to decay exponentially with the distance between qubits in the chain \cite{Lieb1972}. Hence, we define the exponential decay rate $\alpha_q$ by fitting the correlations to an exponential curve,
\begin{equation}
    \braket{Z_q Z_{q+m}}^\text{corr} \sim e^{-\alpha_q m}.
\end{equation}
For each value of $q = 1, \dots, 80$, we calculate $\alpha_q$ by fitting the datapoints $\braket{Z_q Z_{q+m}}^\text{corr}$ for $m = 1, \dots, 20$. Our plotted results compare the decay rates measured using bnZNE and ZNE to the exact values; ultimately, we see more accurate results from bnZNE across the chain. We reject seven high-variance bnZNE datapoints for which $3\sigma > 1.5$ from the fitting process, highlighting a reduced stability in this case.

On the IQM device, we ran 30-qubit \textit{QSim} with between 5 and 40 Trotter layers and present the results in Figure \ref{fig:iqm-combined}. In each case we use the IQM-tailored benchmark circuits (Protocol \ref{protocol:iqm}). In Figure \ref{fig:iqm-combined}(a), we plot the obtained single-qubit expectation value fidelity as a function of the number of Trotter layers $N_T$ after fitting to a quadratic function, which we observed as the consistently best performing fit between linear, quadratic and exponential functions. We ultimately see that bnZNE strongly outperforms standard ZNE even for very deep circuits. In Figure \ref{fig:iqm-combined}(b) we compare the extrapolation step for ZNE and bnZNE, observing how the measured bnZNE noise levels accurately contextualize the relative true noise between different $r$ values to produce a much more accurate extrapolated value.

\section{Conclusion}
We have experimentally demonstrated randomized bnZNE to be a simple, low-cost, and hardware agnostic improvement to one of the most commonly used methods of quantum error mitigation -- zero-noise extrapolation -- which significantly improves the fidelity of expectation values at the fixed cost of running an extra set of benchmark circuits in order to quantify the effect of noise channels on desired observables. We showed how to construct hardware-tailored benchmark circuits for the IBM and IQM hardware with arbitrary applications as well as benchmark circuits tailored to commonly used Floquet style circuits consisting of repetitions of the same core gate layout. We demonstrated the utility of our method by observing significant improvements over standard ZNE for a variety of applications spanning widely different qubit numbers and circuit depths and across two separate hardware platforms. For extremely noise-heavy applications, principally those with high circuit depths, although our method still improves over ZNE, we find that the low signal-to-noise ratio results in a large variance in our fitting in the IBM. This is principally indicative of a hardware application which any given low-cost QEM method would struggle, and is not particular to bnZNE. Furthermore, by exploring a range of different application qubit numbers $N$ and circuit depths $L$, we observed a contradiction to the claim that the simple circuit area metric of $NL$ is an appropriate stand-in for the level expected QEM overhead, instead seeing that circuit depth $L$ is a much stronger proxy for expected bnZNE and ZNE performance. 

We expect that -- due to its simplicity of implementation, low overhead and demonstrated performance -- our bnZNE method will be of interest to users of NISQ and early-fault tolerant devices as a clear improvement over the already commonly-used ZNE, with its range of utility increasing as hardware error rates continue to improve.

\section{Acknowledgments; data and code availability}

JH and PS acknowledge funding from the DLR Quantum Computing Initiative \cite{ALQU} and the German Federal Ministry of Research, Technology and Space \cite{BMFTR}. KL acknowledges funding from  the Q-Neko project, which has received funding from the European Union's Horizon Europe research and innovation programme under Grant Agreement No. 101241875.

Some program code and data produced during this work can be found in the associated GitHub repository \cite{Github}.

We acknowledge the use of IBM Quantum services for this work. The views expressed are those of the authors, and do not reflect the official policy or position of IBM or the IBM Quantum team. All quantum circuits were produced using Qiskit \cite{qiskit2024}. We acknowledge IQM for providing access to their quantum computing resources through the DLR ALQU project.

\bibliographystyle{IEEEtran}
\bibliography{references}


\end{document}